\newif\iflong

\longtrue
\iflong
\documentclass[onecolumn,10pt]{IEEEtran}
\else
\documentclass[conference,10pt]{IEEEtran}
\fi

\usepackage{times}
\usepackage{xr}

\usepackage{amsthm}
\usepackage{amsmath}    
\usepackage{amssymb}    
\usepackage{mathrsfs}   
\usepackage{epsfig}     
\usepackage{graphicx}
\usepackage{url}

\usepackage{url}
\usepackage{color}
\usepackage{booktabs}

\usepackage{subfigure}

\usepackage{amsfonts}
\usepackage{dsfont}
\usepackage{url}
\usepackage{textcomp}
\usepackage{multirow}
\usepackage[numbers,sort&compress]{natbib}

\usepackage{enumitem}

\usepackage[hmargin=1.3cm,vmargin=1.3cm]{geometry}

\graphicspath{{}{figs/}{../}{../figs/}{../Monograph/monograph/nowformat/}}

\newif\ifmonograph
\monographfalse

\renewcommand{\caption}[1]{\singlespacing\hangcaption{#1}\normalspacing}

\usepackage{cleveref}

\usepackage{amsmath}
\usepackage{amsfonts}
\usepackage{dsfont}
\usepackage{mathrsfs}

\usepackage{multirow}

\usepackage{algorithm}

\usepackage[noend]{algcompatible}

\newcommand{\ep}{\epsilon}

\newcommand{\C}{{\mathcal C}}

\newcommand{\G}{{\mathcal G}}

\newcommand{\E}{{\mathcal E}}
\newcommand{\V}{{\mathcal V}}

\newcommand{\U}{{\mathcal U}}
\newcommand{\T}{{\mathcal T}}

\newcommand{\Rs}{{\mathcal R}}

\newcommand{\s}{{\bf s}}

\newcommand{\st}{{\rm st}}

\newcommand{\bA}{\mathsf{A}}
\newcommand{\bC}{\mathsf{C}}
\newcommand{\bG}{\mathsf{G}}
\newcommand{\bT}{\mathsf{T}}

\newcommand{\sC}{\overline{\mathcal C}}

\newcommand{\comment}[1]{{}}

\newcommand{\g}{{G}}

\renewcommand{\r}{{\bf r}}
\newcommand{\x}{{\bf x}}
\newcommand{\y}{{\bf y}}

\newcommand{\din}{d_{\rm in}}
\newcommand{\dout}{d_{\rm out}}

\newcommand{\lcrit}{\ell_{\rm crit}}

\newcommand{\defi}{\triangleq}

\newcommand{\la}{\phi}
\newcommand{\ch}{\la}

\newcommand{\ec}{{\rm ec}}

\newcommand{\concat}{\oplus}

\newcounter{constcount}
\newcounter{numcount}
\newcounter{thmcnt}
  \let\Oldsection\section
  
\renewcommand{\section}{\stepcounter{thmcnt}\Oldsection}

\allowdisplaybreaks

\newtheorem{theorem}{Theorem} 
\newtheorem{lemma}{Lemma} 
\newtheorem{definition}{Definition} 
\newtheorem{claim}{Claim} 
\newtheorem{cor}{Corollary} 

\newtheorem{remark}{Remark} 
\newtheorem{question}{Question} 

\newcounter{examplecounter}
\newcommand{\aln}[1]{\begin{align*}#1\end{align*}}

\newcommand{\al}[1]{\begin{align}#1\end{align}}

\def\Item$#1${\item $\displaystyle#1$
   \hfill\refstepcounter{equation}(\theequation)}

\newcommand{\bea}{\begin{eqnarray}}
\newcommand{\eea}{\end{eqnarray}}
\newcommand{\beas}{\begin{eqnarray*}}
\newcommand{\eeas}{\end{eqnarray*}}

\begin{document}

\title{Partial DNA Assembly: A Rate-Distortion Perspective}


\author{
Ilan Shomorony$^{1}$, Govinda M. Kamath$^{2}$, Fei Xia$^{3}$, Thomas A. Courtade$^{1}$, and David N. Tse$^{2}$ \\
$^{1}$ University of California, Berkeley,  USA, $^{2}$ Stanford University, Stanford, USA,
$^{3}$ Tsinghua University, China.\\
Email: ilan.shomorony@berkeley.edu, gkamath@stanford.edu, xf12@mails.tsinghua.edu.cn,\\ courtade@berkeley.edu, dntse@stanford.edu.
\thanks{ \scriptsize This work is partially supported by the Center for Science of Information (CSoI), an NSF
Science and Technology Center, under grant agreement CCF-0939370.}}


\maketitle

\begin{abstract}
Earlier formulations of the DNA assembly problem were all in the context of perfect assembly; i.e., given a set of reads from a long genome sequence, is it possible to perfectly reconstruct the original sequence?
In practice, however, it is very often the case that the read data is not sufficiently rich to permit unambiguous reconstruction of the original sequence.
While a natural generalization of the perfect assembly formulation to these cases would be to consider a rate-distortion framework, partial assemblies are usually represented in terms of an \emph{assembly graph}, making the definition of a distortion measure challenging.
In this work, we introduce a distortion function for assembly graphs that can be understood as the logarithm of the number of Eulerian cycles in the assembly graph, each of which correspond to a candidate assembly that could have generated the observed reads.
We also introduce an algorithm for the construction of an assembly graph and analyze its performance on real genomes.
\end{abstract}

%

\section{Introduction}
\label{sec:intro}
\iflong

The cost of DNA sequencing has been falling at a rate exceeding Moore's law. 
The dominant technology, called  \emph{shotgun sequencing} involves obtaining a large number of 
fragments called reads from random locations on the DNA sequence.
This technology comes in two flavors:
\begin{enumerate}
 \item \emph{Short-read technologies}, which generate reads typically shorter than
 $200$ base pairs (bp), with error rates around $1\%$, with substitutions being the primary form of 
 errors.
 \item \emph{Long-read technologies}, which generate reads of length around $10,000$ bp, with error rates around $15\%$, with insertions and deletions being the primary form of 
 errors.
\end{enumerate}

The reads obtained from either technology are then merged to each other based
on regions of overlap using an \emph{assembly algorithm} 
to obtain an estimate of the DNA sequence. 
During the last decade, several such algorithms were developed first aimed at short-read sequencing technologies and, more recently, focused on long-read technologies.
While these approaches  attained 
varied degrees of success in the assembly of many genomes, 
very few of them are known to provide any kind of performance guarantee.

A theoretical framework 
to assess the performance of various algorithms relative to the \emph{fundamental limits} for DNA assembly was proposed in \cite{BBT}.
However, this framework focuses on \emph{perfect assembly}; i.e., 
when the goal is to reconstruct the whole genome perfectly.
In particular, a critical read length $\lcrit$, defined as a function of the repeat patterns of a given genome \cite{Ukkonen,BBT}, is proved to be a fundamental lower bound for perfect assembly, and shown to be achievable by their proposed algorithm.
Nonetheless, for real genomes, $\lcrit$ can be very large, and 
read data for many practical DNA sequencing 
projects does not meet the information-theoretic lower bounds from
\cite{Ukkonen,BBT}, 
rendering the task of perfect assembly fundamentally impossible.
 This makes the derivation
 of a theoretic framework to compare algorithms in terms of partial assembly of paramount importance.

 The first step in constructing a theoretical framework for partial assembly is to 
 select an appropriate measure of the quality of a partial assembly.
 Measuring the quality of partial assemblies is a challenging problem. In practice,
 a metric that is often used is the N50. 
 To describe N50, recall that a \emph{contig} is an unambiguous sequence in 
 a genome that an assembly returns. The N50 of an assembly is then defined as the largest 
 length $\ell$ such that the sum of the lengths of contigs at least $\ell$ long accounts for at
 least half of the sum of the lengths of all contigs returned. 
 This is practically a very
 popular metric, because it does not require knowledge of the ground truth genome
 (which is usually not known in practice) to compute. 
 However, the fact that N50 does not depend upon the ground truth genome makes it
 mathematically unsatisfying. For example,  an algorithm could just output a random string over 
 $\Sigma= \{ \bA,\bC,\bG,\bT\}$ of length $1$ trillion and obtain an N50 of $1$ trillion,
 despite the fact that the output is unrelated to the target genome.
 Another discouraging aspect of N50 is that it cannot capture
 what is known about the relative position between the contigs. 
 In particular, contigs are typically extracted from an \emph{assembly graph}, which is basically a graph with
 the contigs as vertices, and an edge from contig $u$ to contig $v$ if 
 contig $v$ comes after one copy of contig $u$. N50
 does not account for any of the structural information contained in such a graph.

   In this manuscript, we introduce a distortion metric that attempts to capture how good an assembly graph is.
   Roughly speaking, this measure coincides with the logarithm of the number of Eulerian cycles in the assembly graph.
   Intuitively, every Eulerian cycle corresponds to a distinct assembly and all of them explain the data equally well; as such, the distortion represents the missing information still needed for perfect assembly.  
    To the best of our knowledge, this is the first work in this direction.

%

With the yardstick defined, 
we then seek an assembly algorithm whose performance can be characterized in terms of the proposed distortion measure.
While \emph{de Bruijn} graph-based algorithms \cite{PevznerEuler,IDBA,Velvet,Spades}
are better understood from a theoretical standpoint \cite{PevznerEuler,BBT},
and would constitute better candidates for the distortion analysis, they are
not very relevant in the context of assembly from long-read technologies.
This is due to their high sensitivity to read errors, which prevents them from leveraging the potential of long-read
technologies (all of which have error rates above $10\%$ and will 
 continue to have for the foreseeable 
  future \cite[Sections $1,3$]{DAligner}).
 In contrast, overlap-based assembly approaches (e.g., string graphs \cite{MyersString}) are better suited to long-read, high-error sequencing.   
 This class of algorithms relies on the identification of long overlaps between reads, which is inherently robust to errors \cite{DAligner},
and has been recently shown to attain the same theoretical performance as de Bruijn graph-based approaches in terms
of perfect assembly \cite{NSG}.
 In this work, we propose a new overlap-based assembly algorithm
 and introduce techniques to provide theoretical guarantees in terms of the new distortion measure.
 
%

\fi
 
\section{Problem Setting} \label{sec:setting}

Let $\x$ be a string of $\ell$ symbols from the alphabet $\Sigma = \{\bA,\bC,\bG,\bT\}$.
We let $|\x| = n$ be the length of the string, and $\x[i]$ be its $i$th symbol.
A substring of $\x$ is a contiguous interval of the symbols in $\x$, and is denoted as 
$\x[i:j] \triangleq (\x[i],\x[i+1],...,\x[j])$.
A substring of the form $\x[1:\ell]$ is called a prefix (or an $\ell$-prefix) of $\x$.
Similarly, a substring of the form $\x[|\x|-\ell+1:|\x|]$ is a suffix (or an $\ell$-suffix) of $\x$. 
We say that strings $\x$ and $\y$ have an \emph{overlap} of length $\ell$ if the $\ell$-suffix of $\x$ and the $\ell$-prefix of $\y$ are equal. 
We let $\x \concat \y$ denote the concatenation of $\x$ and $\y$.

We assume that there exists an unknown target DNA sequence $\s$ of length $|\s| = \g$ which we wish to assemble from a set of $N$ reads $\Rs$.
Throughout the paper, we will make two simplifying assumptions about the set of reads: 
\begin{enumerate}[label=(A\arabic*),leftmargin=9mm]
\item All reads in $\Rs$ have length $L$.
\item The reads in $\Rs$ are error-free.
\end{enumerate}
The first assumption is made to simplify the exposition of the results.
The second assumption is motivated by the existence of overlapping tools (such 
as DAligner \cite{DAligner}), which can efficiently 
identify significant matches between reads at error rates of 
over $15\%$.
The algorithm described in this manuscript can be adapted to work with the approximate matches found by such tools, essentially by treating them  as exact matches.


For ease of exposition, we will assume that $\s$ is a circular sequence of length $\g$; i.e., $\s[t+\g] = \s[t]$ for any $t$.
This way we will avoid edge effects and a read $\mathbf{x} \in \Rs$ can correspond to any substring $\s[t:t+L-1]$, for $t=1,...,\g$.
We will use the standard Poisson sampling model for shotgun sequencing.
This means that each of the $N$ reads is drawn independently and uniformly at random from the set of length-$L$ substrings of $\s$, $\{\s[t:t+L-1] : t = 1,...,\g\}$.

\subsection{Repeats and Bridging}

A \emph{repeat} of length $\ell$ in $\s$ is a substring $\x \in \Sigma^\ell$ appearing at distinct positions $t_1$ and $t_2$ in $\s$; i.e., $\s[t_1:t_1+\ell-1] = \s[ t_2: t_2+\ell-1] = \x$, that is maximal; i.e., $\s[t_1-1] \ne \s[t_2-1]$ and $\s[t_1+\ell] \ne \s[t_2+\ell]$.
A repeat is bridged if there is a read that extends beyond one copy of the repeat in both directions, as shown in Fig.~\ref{fig:bridgedrep}.
A repeat is doubly-bridged if both copies are bridged.
\iflong
\begin{figure}[ht] 
	\center
	\vspace{-3mm}
       \includegraphics[width=0.5\linewidth]{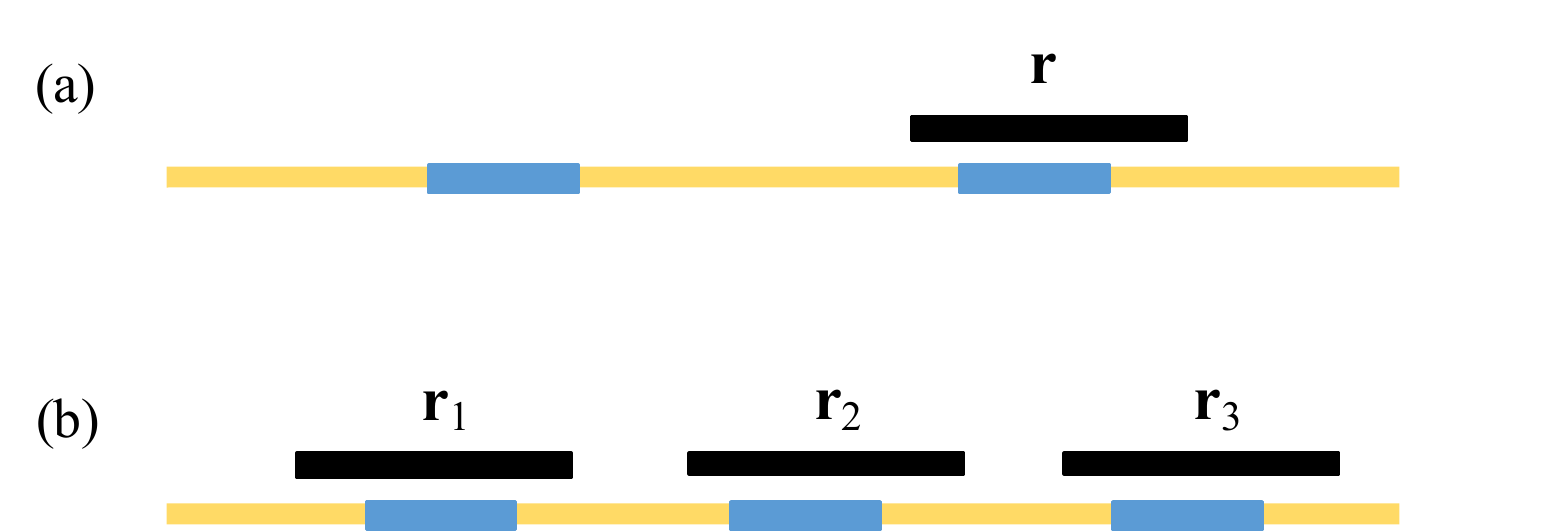} 
       \caption{(a) Illustration of a bridged repeat in $\s$; (b) Illustration of a triple repeat all-bridged by reads $r_1$, $r_2$ and $r_3$.\label{fig:bridgedrep}}
\end{figure}
\fi
Similarly, a triple repeat of length $\ell$ is a substring $\x$ that appears at three distinct locations in $\s$ (possibly overlapping); i.e.,
$\s[t_1:t_1+\ell-1]=\s[t_2:t_2+\ell-1]=\s[t_3:t_3+\ell-1] = \x$ for distinct $t_1$, $t_2$ and $t_3$ (modulo $G$, given the circular DNA assumption), 
and is maximal (that is, all three copies can not be extended in a direction).
A triple repeat is said to be bridged if at least one of its copies is bridged.
It is said to be \emph{all-bridged} if all of its copies are bridged, as illustrated in Fig.~\ref{fig:bridgedrep}(b), and \emph{all-unbridged} if none of its copies are bridged.
%

\section{A Distortion Metric for Assembly Graphs}\label{sec:distmetric}


To motivate our notion of distortion for assembly graphs, let us first consider 
an idealized setting in which \emph{all} reads of length $L$ from $\s$ are given.
We call this ensemble of reads the $L$-mer composition of $\s$,
defined as the multiset
\al{
\C_L(\s) = \left\{ \s[i:i+L-1] : 1 \leq i \leq G \right\}.
}
We will let $\sC_L(\s)$ represent the support of $\C_L(\s)$; i.e., $\C_L(\s)$ without the copy counts.
The $L$-mer composition $\C_L(\s)$ does not, in general, determine $\s$ unambiguously.  
In particular, any sequence $\x$ with  $\C_L(\x)=\C_L(\s)$ is indistinguishable from
$\s$ when only $\C_L(\s)$ is observed.  
Thus, one requires at least
\al{
\log \left| \left\{ \x : \C_L(\x) = \C_L(\s) \right\} \right|
}
additional bits to determine $\s$ unambiguously from all other sequences having the same $L$-mer composition.
This uncertainty characterization can be translated to the language of sequence graphs through the notion of a $k$-mer graph\footnote{ 
In the assembly literature, such graphs are sometimes referred to as de Bruijn graphs.
However, since our proposed algorithm is not a de Bruijn graph based algorithm in the usual sense of \cite{PevznerEuler}, we avoid the terminology. 
}
of $\s$, $B_k(\s)$ \cite{PevznerEuler}.
\begin{definition}
The multigraph $B_k(\s)$ has $\sC_{k-1}(\s)$ as its node set and, for $\x,\y \in \sC_{k-1}(\s)$, we place $m$ edges from $\x$ to $\y$ if the $k$-mer $\x \concat \y[k-1]$ has multiplicity $m$ in $\C_k(\s)$. 
\end{definition}
It is easy to see that $B_k(\s)$ is an Eulerian graph for any $k$, and every $\x$ with $\C_k(\x) = \C_k(\s)$ corresponds to a distinct Eulerian cycle in $B_k(\s)$.
Furthermore, two Eulerian cycles that are distinct up to edge multiplicities correspond to distinct sequences.
Therefore, a natural measure of the ``distortion'' of $B_k(\s)$ as an assembly graph of $\s$ would be
\al{ \label{eq:kdist}
D_k(\s) \defi \log \ec(B_k(\s)),
}
where $\ec(G)$ is the number of Eulerian cycles in $G$, distinct up to edge multiplicity.
We point out that the number of Eulerian cycles in a $k$-mer graph has been previously used in the related context of DNA-based storage channels \cite{OlgicaDNACodes}.

\begin{figure}[h]
\centering
\includegraphics[width = 0.5\linewidth]{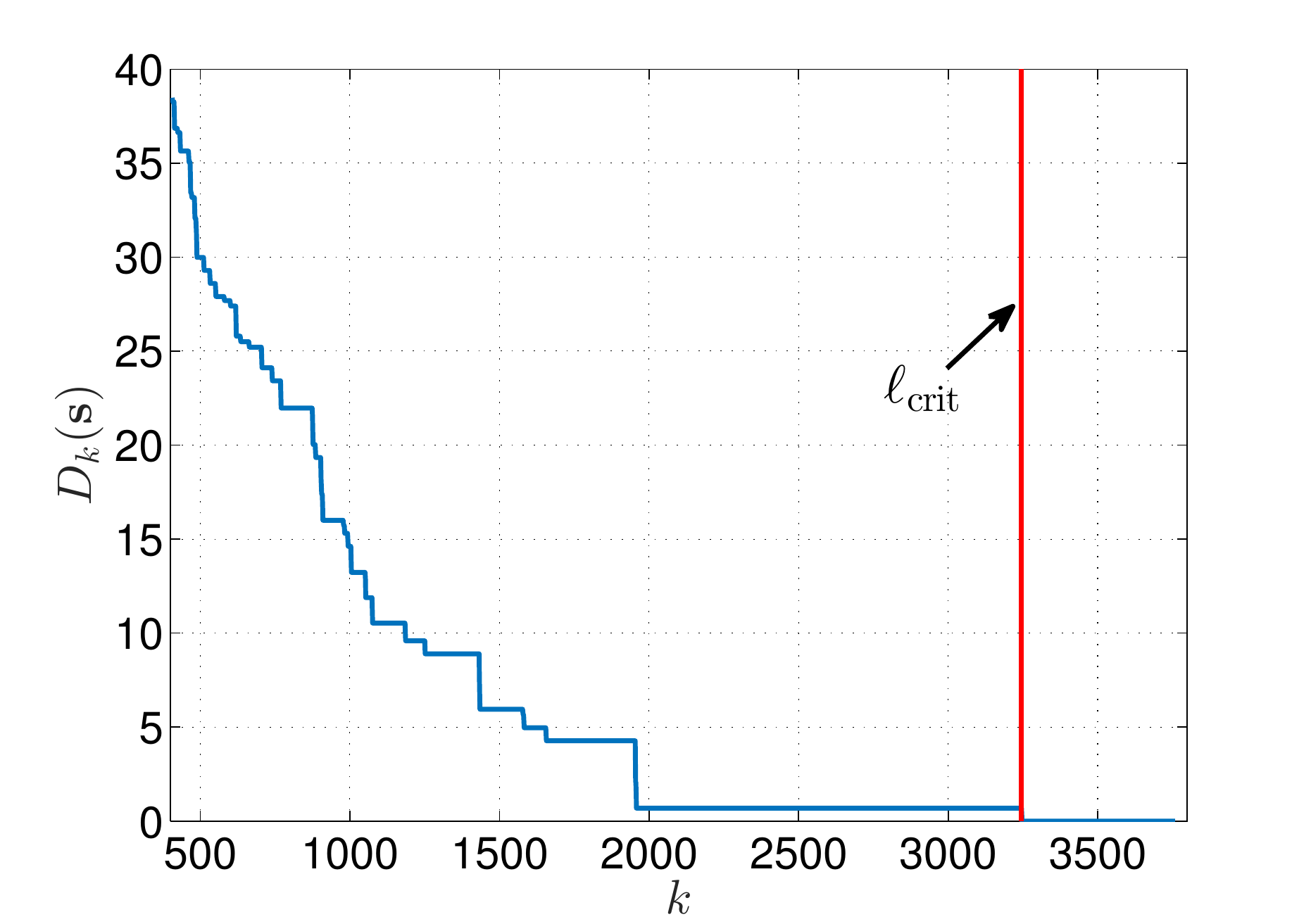}
\caption{$D_k(\s)$ as a function of $k$, when $\s$ is the genome of \emph{E. coli 536}. Notice that $D_k(\s)$ reaches zero when $k = \lcrit(\s)$ \cite{BBT}.} 
\label{fig:kdist}
\end{figure}

As illustrated in Fig.~\ref{fig:kdist},
$D_k(\s)$ can be computed for real genomes, and can be interpreted as a lower bound on how good an assembly from reads of length $k$ can be.


In the actual setting for the assembly problem, however, one does not have access to the entire $L$-mer composition of $\s$, nor can be expected to perfectly construct $B_k(\s)$ for some $k < L$ (other than for small values of $k$).
Hence, when defining a distortion metric for assembly graphs, one must consider a larger class of graphs than $B_k(\s)$.
In this work, we will consider the following:

\begin{definition}\label{def:seq_graph}
A sequence graph $G = (V,E,\ch)$ of order $k$ is a directed multigraph where each edge $e \in E$ is labeled with a $k$-mer $\ch(e) \in \Sigma^k$, and each node $v \in V$ is labeled with a $(k-1)$-mer $\ch(v) \in \Sigma^{k-1}$ satisfying the property that if $\ch(u,v) = \x$, then $\ch(u) = \x[1:k-1]$ and $\ch(v) = \x[2:k]$.
\end{definition}

Notice that any path $p = (v_1,...,v_\ell)$ on a sequence graph of order $k$ naturally defines a length-$(\ell-2+k)$ string 
\aln{
\st(p) \defi \ch(v_1,v_2)[1]\concat ... \concat \ch(v_{\ell - 2},v_{\ell-1})[1] \concat \ch(v_{\ell - 1},v_\ell).}
If a path $p = (v_1,...,v_\ell)$ ends in a node with out-degree zero, it will be called a graph suffix, and if it starts in a node with in-degree zero, it will be called a graph prefix.

\begin{definition}
A Chinese Postman cycle in a sequence graph $G$, is a cycle that traverses every edge at least once.
\end{definition}

A natural formulation for the genome assembly problem is to identify a Chinese Postman cycle in the constructed sequence graph which corresponds to the true sequence \cite{Nagarajan,Medvedev}.

%

\begin{definition} \label{def:suff}
A sequence graph $G$ is said to be sufficient (for the assembly of $\s$) if it contains a Chinese Postman cycle $c_\s$ such that $\st(c_\s) = \s$ (up to cyclic shifts).
\end{definition}

While it is natural to define the goal of the partial assembly problem to be the construction of a sufficient sequence graph,  it is typically unreasonable to expect the assembly algorithm to correctly estimate the multiplicities of all the edges; i.e., the number of times $c_\s$ traverses each edge.
The reason is that the length of the genome is not known in advance, and hence neither is the coverage depth (i.e., the average number of reads covering a given position).
Other practical issues like uneven coverage and sequence specific biases only add to the difficulty there.
Thus our distortion metric should not penalize incorrect multiplicities, and thus not require the produced graph to be Eulerian.
To define our distortion metric, we will consider an ``Eulerian version'' of the constructed sequence graph.
More precisely, if $G = (V,E,\ch)$ is a sufficient sequence graph and $c_\s$ is a Chinese Postman cycle in $G$ corresponding to the sequence $\s$,
we will let $G[{\s}]$ be the multigraph obtained by setting the multiplicity of edge $e$ to be the number of times $c_\s$ traverses $e$.

%
%
%
%
%
%
%
%
%
%
%
%
%
%
%
%


\begin{definition}  \label{def:dist}
We define the distortion of a sequence graph $G$ as
\al{ \label{eq:dist}
D(G,\s) \defi \left\{ 
\begin{array}{ll}
\log \ec(G[\s]) & \text{if $G$ is a sufficient} \\& \text{sequence graph for $\s$} \\
D_1(\s) +1& \text{otherwise} 
\end{array} 
\right. 
}
where $\ec(G)$ is the number of Eulerian cycles in $G$ that are distinct up to edge multiplicities,
and $D_1(\s)$ is the distortion achieved by the $1$-mer graph of $\s$, $B_{1}(\s)$.
%
\end{definition}

\iflong
\begin{figure}[b]
\centering
\vspace{-4mm}
\includegraphics[width = 0.5\linewidth]{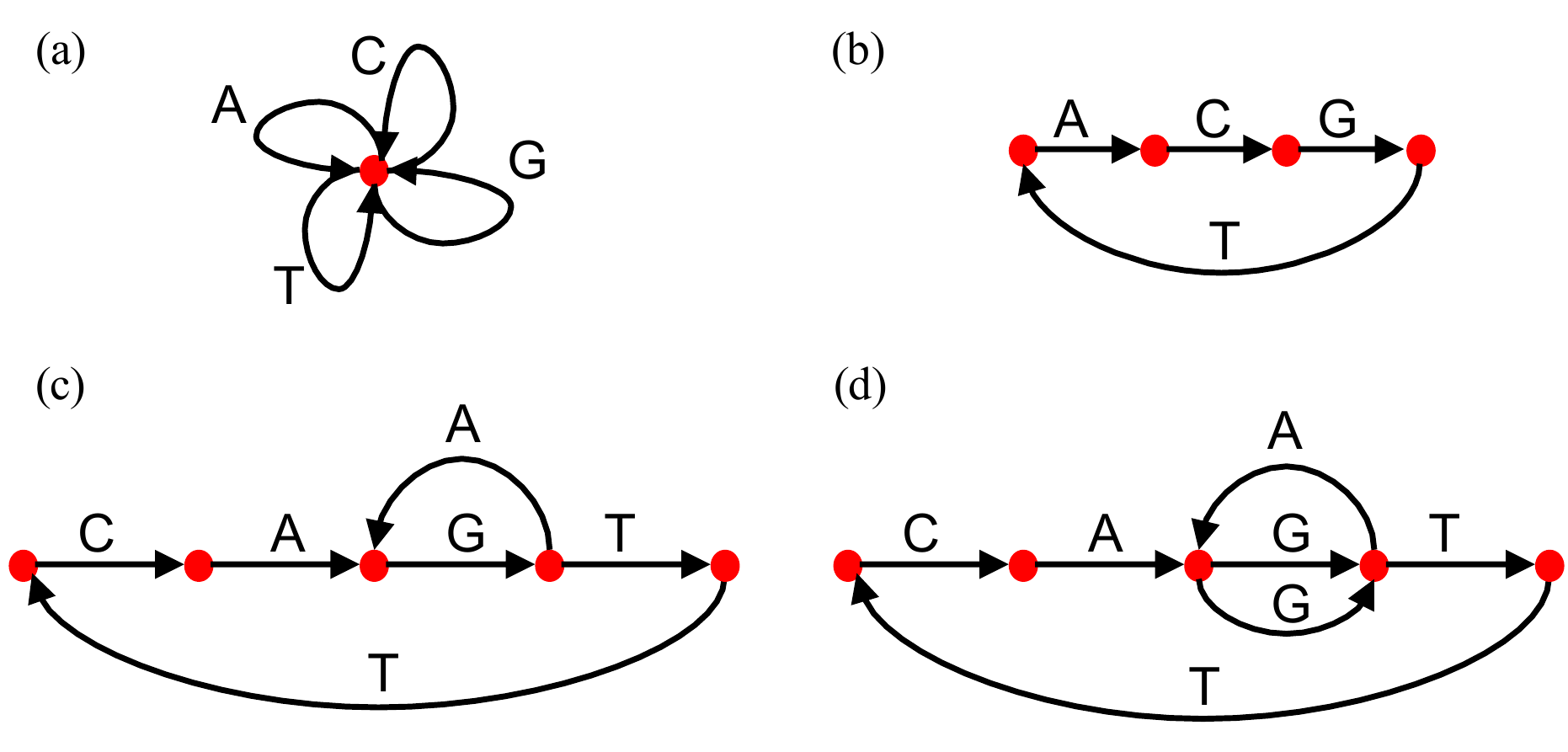}
\caption{(a) The trivial sequence graph $G_0$ is always sufficient. (b,c,d) An 
example of the distortion computed for the assembly of a cyclic sequence 
$\s = \mathsf{C\,A\,G\,A\,G\,T\,T}$ is shown.
If the graph in (b) is returned by an assembly algorithm, then as 
the graph is not a sufficient sequence graph with respect to $\s$,
the distortion is computed to be $\log \left[ \frac{1}{7} \binom{7}{2,1,2,2}\right]+1 =2.95$.
If the sequence graph $G$ of order $k=1$ in (c) is returned, then $G[\s]$ is as shown in (d). 
The distortion of this graph is $0$ as there
is exactly one Eulerian cycle there (modulo differences in traversing edges between the same two vertices).
} 
\label{fig:example}
\end{figure}
\fi

We note that if $\s$ contains all of $\{\bA,\bC,\bG,\bT\}$, then $D_1(\s)$ would be the distortion achieved by
the graph $G_0$, shown in Figure \ref{fig:example}(a).
\iflong
Notice that, when $G$ is not sufficient, we set the distortion to be worse than the distortion $D_1(\s)$. 
\fi
It is not difficult to see that
the distortion of any sufficient sequence graph is at most $D_1(\s)$. 
\iflong
The definition makes sure that
the distortion achieved by any sufficient sequence graph is less than any graph that 
is not sufficient. 
\fi
Fig.~\ref{fig:example} shows the computation of this distortion in a toy example.


%

\iflong

\section{A Greedy Algorithm for Partial Assembly and Associated Guarantees} \label{sec:algorithm}

\fi


In this section we describe an algorithm to assemble a sequence graph.
We then analyze its performance in terms of its ability to produce a  sufficient sequence graph and the resulting distortion.

The algorithm can be seen as a generalization of the greedy algorithm for sequence assembly \cite{UkkonenGreedy}.
In the standard greedy algorithm, prefixes and suffixes of reads are iteratively merged in order to produce a single sequence.
However, when an incorrect merging occurs, it has no way of detecting and fixing it at later iterations.
Our algorithm overcomes this issue by allowing 
%
%
a read prefix/suffix to be merged to the \emph{interior} of another read, or to a previously merged prefix/suffix,
as illustrated in \iflong Fig.~\ref{fig:merging} \fi.
As we will show, this additional flexibility is helpful in constructing a sufficient sequence graph in the sense of Definition~\ref{def:suff}, making the algorithm robust from the point of view of partial assembly.

\iflong
\begin{figure}[h]
\centering
\includegraphics[width = 0.5\linewidth]{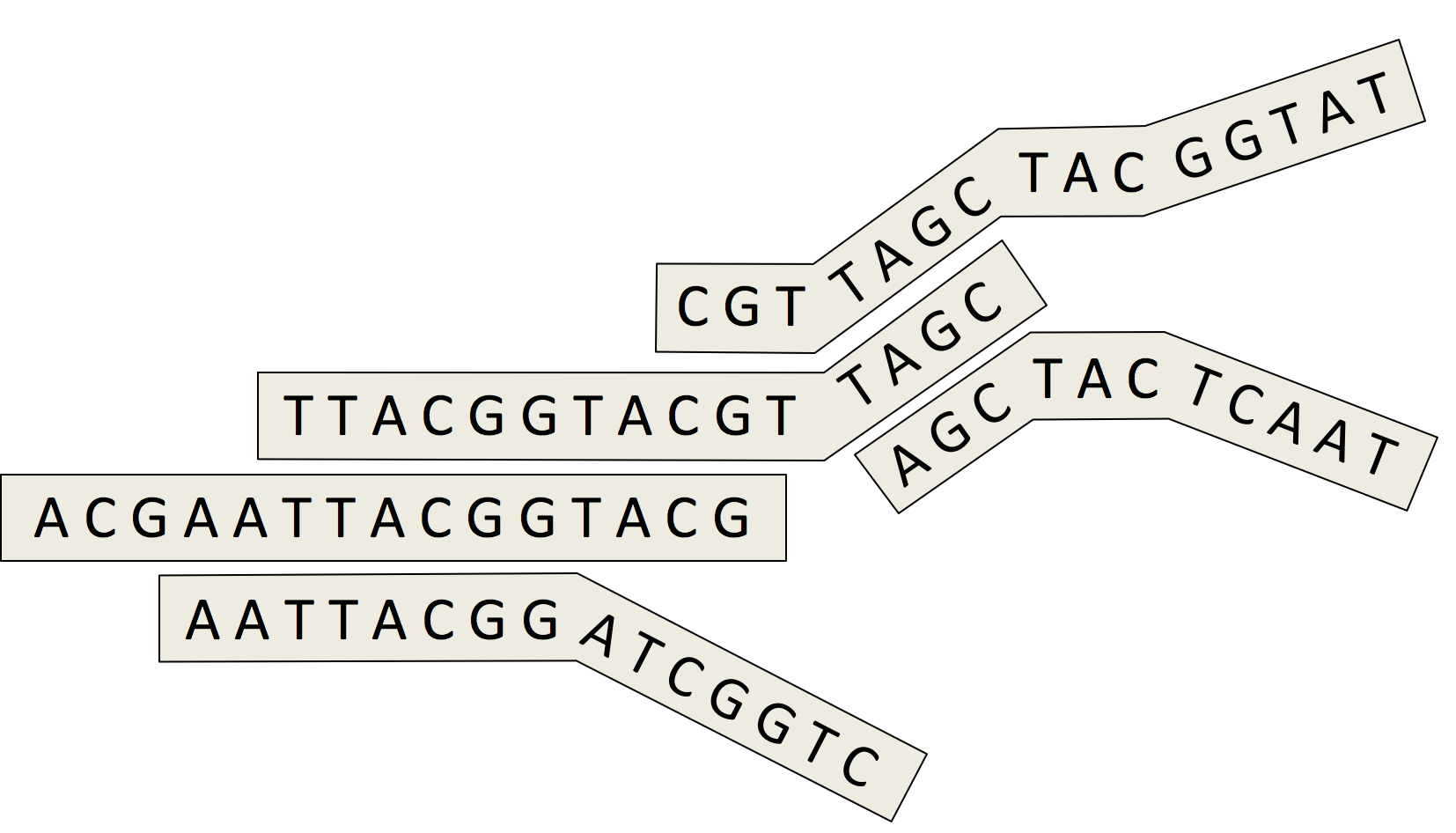}
\caption{In the greedy merging algorithm, we allow matches between a prefix/suffix of a read and the interior of another read, producing a graph that is not a line, as is the case with the standard greedy algorithm \cite{UkkonenGreedy}} 
\label{fig:merging}
\end{figure}
\fi

Our algorithm will maintain at all times a sequence graph in the sense of Definition~\ref{def:seq_graph}, where each read $\r_i \in \Rs$ corresponds to a path $p_i$ with $L-k+2$ nodes and $L-k+1$ edges, which correspond to the $L-k+1$ consecutive $k$-mers of $\r_i$.
Initially, all $N$ paths will be disjoint components of the graph, as illustrated in 
\iflong Fig.~\ref{fig:initial}. \fi
\iflong
\begin{figure}[h]
\centering
\includegraphics[width = 0.5\linewidth]{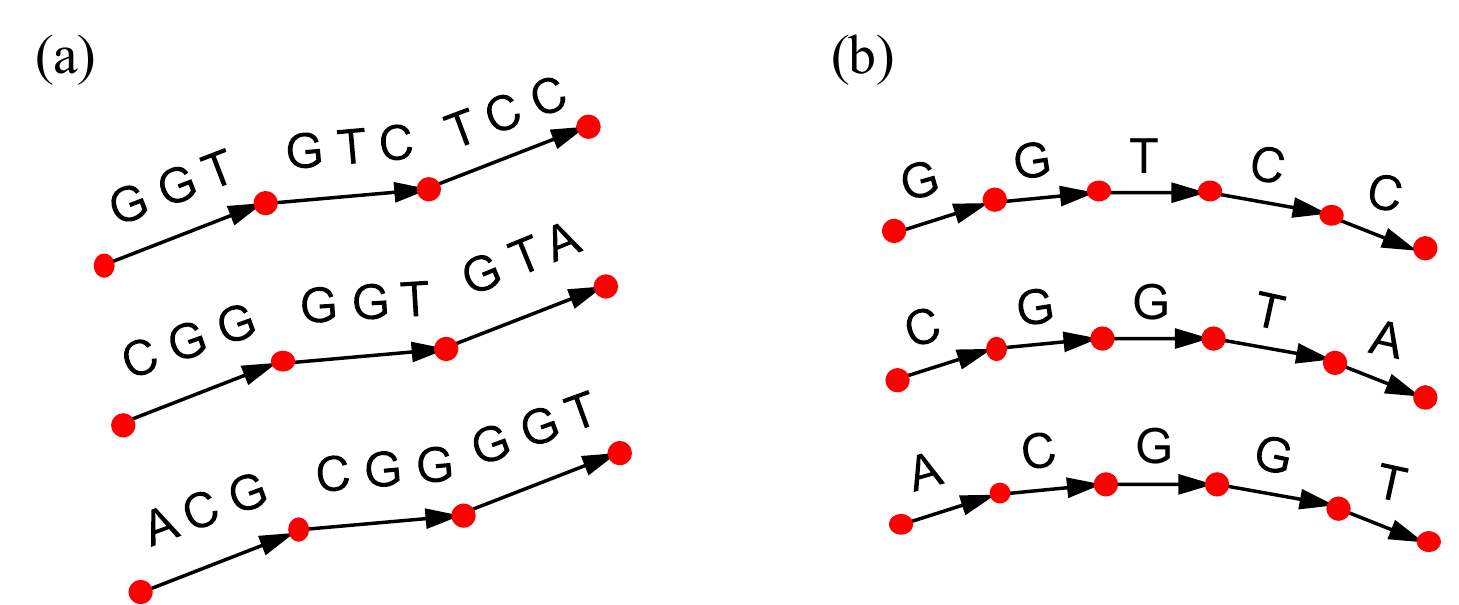}
\caption{Initial sequence graph for reads $\mathsf{G\,G\,T\,C\,C}$, $\mathsf{C\,G\,G\,T\,A}$, and $\mathsf{A\,C\,G\,G\,T}$ for (a) $k=3$  and (b) $k=1$. Notice that a match of $\ell$ symbols between two reads corresponds to a path of $\ell - k + 1$ edges. We point out that when $k=1$  the nodes do not have a label associated with them.}
\label{fig:initial}
\end{figure}
\fi
The algorithm then proceeds by finding matches between a previously unused prefix or suffix and any part of another read, and merging the corresponding paths.
The algorithm is termed greedy since it searches for matches in decreasing order of length.


\begin{algorithm}[htb]  
\caption{Greedy merging algorithm}  \label{alg:greedyalg}
\begin{algorithmic}[1]
\STATE Input: Initial sequence graph (\iflong see Fig.~\ref{fig:initial} \fi), and parameter $k$
\FOR {$\ell = L,L-1,L-2,...,k$}
	\STATE \mbox{$X \gets {{\{}} \x \in \Sigma^\ell :  \text{$\x$ is a current graph prefix or suffix that}$} 
	
	\quad \quad appears in more than one read{{\}}}
	\FOR {$\x \in X$}
		\STATE {Merge the path corresponding to $\x$ from all reads that
		
		  contain the substring $\x$}  \label{line:merge}
	\ENDFOR
\ENDFOR
\STATE Output: Resulting sequence graph of order $k$  
\end{algorithmic}
\end{algorithm}

\iflong

The parameter $k$ should be chosen as the minimum overlap we expect adjacent reads to have, and can be made large for sequencing experiments with high coverage depth.
For instance, when assembling long reads ($10,000$ bp) with high error rates, a typical choice for the minimum overlap is $1000$ \cite{DAligner}.
We proceed to analyze the distortion achieved by the sequence graph that Algorithm \ref{alg:greedyalg}
outputs in two steps:
\begin{itemize}
 \item We first obtain conditions for the sequence graph to be sufficient. 
 In other words, we obtain conditions under which the the distortion 
 achieved by the sequence graph is in the case $1$ of the distortion in Definition \ref{def:dist}.
 \item Then we characterize conditions under which the distortion of the resulting sufficient sequence graph can be upper bounded by $D_k(\s)$ for some $k > 1$.
\end{itemize}

\fi


\begin{definition}\label{def:kcovers}
We say that $\Rs$ $k$-covers the sequence $\s$ if there is a read starting in every $k$-length substring of $\s$.
\end{definition}


\begin{theorem} \label{thm:correctness}
Algorithm \ref{alg:greedyalg} constructs a sufficient sequence graph of order $k$ if the set of reads $\Rs$ 
$k$-covers the sequence $\s$ and every 
 triple repeat is either unbridged or all-bridged.
\end{theorem}

As described in Section \ref{sec:numerical}, 
given the conditions in Theorem~\ref{thm:correctness}, one can bound the probability that the graph produced by Algorithm \ref{alg:greedyalg} is not sufficient.
This bound can then be translated into a value of coverage depth $c = NL/G$ for which the resulting sequence graph 
is sufficient with a desired probability $1-\ep$.
This is illustrated in Fig.~\ref{fig:saureus_sufficient} for the \emph{S. aureus} genome from the GAGE dataset \cite{GAGE}.
We notice that for values of $L$ that are far from the length of some triple repeat, a small coverage depth suffices.


We remark that an interesting open question is to determine if the non-monotonicity caused by the peaks in required coverage near triple repeat lengths (as shown in Fig.~\ref{fig:saureus_sufficient})
 represents a fundamental barrier or a limitation of the algorithm.
 Most existing algorithms face challenges when there are triple repeats
 of lengths that are close to the read length. In fact, overlap-based algorithms also suffer from 
 similar problems when there are double repeats of lengths close to $L$.

\iflong
\begin{figure}[b]
\centering
\vspace{-4mm}
\includegraphics[width = 0.5\linewidth,trim={1.9cm 6.3cm 2cm 6.3cm},clip]{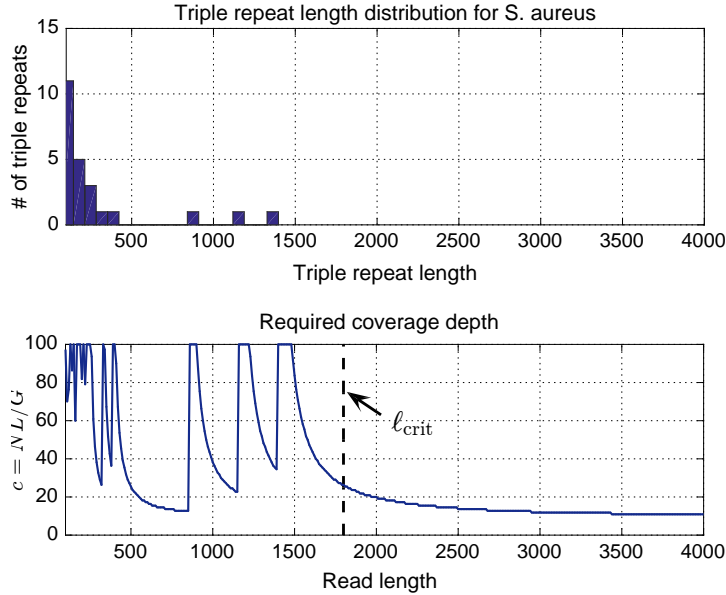}
\caption{Distribution of triple repeat lengths on \emph{S. aureus}, and coverage depth required for the conditions in Theorem~\ref{thm:correctness} to be achieved with probability $0.99$.} 
\label{fig:saureus_sufficient}
\end{figure}
\fi


In addition to the sufficiency property guaranteed by Theorem~\ref{thm:correctness}, we need a way to characterize the distortion achieved by the resulting graph. 
To do so, we will bound the distortion achieved by assembling reads of length $L$ by the quantity $D_q(\s)$, defined in (\ref{eq:kdist}), for some $q < L$.
We begin with a definition.

\begin{definition}
Two repeats $\s[a_1: a_1+\ell]$, $\s[a_2: a_2+\ell]$ and $\s[b_1: b_1+m]$, $\s[b_2: b_2+m]$ are said to be linked if $a_2 < b_1 \leq a_2 + \ell + 1$.
We call $a_2 + \ell + 1-b_1$ the link length.
\end{definition}

\iflong
The importance of introducing linked repeats is that, as illustrated in Fig.~\ref{fig:linked}, they are potential causes of ambiguity in the sequence graph.
\fi

\begin{theorem} \label{thm:distortion}
Suppose that the set of reads $\Rs$ from the sequence $\s$ satisfies the following conditions:
\begin{enumerate}
\item each triple repeat is either all-bridged or all-unbridged,
\item all repeats of length $\leq q$ are doubly-bridged,
\item for all pairs of linked repeats with link length $\ell$ satisfying $k-1 \leq \ell \leq q$, at least one is doubly-bridged.
\end{enumerate}
Then the sufficient sequence graph $G$ produced by Algorithm \ref{alg:greedyalg} has a distortion satisfying
\iflong
\aln{
D(G,\s) \leq D_q(\s).
}
\fi
\end{theorem}

\iflong
\begin{figure}[t]
\centering
\includegraphics[width = 0.5\linewidth]{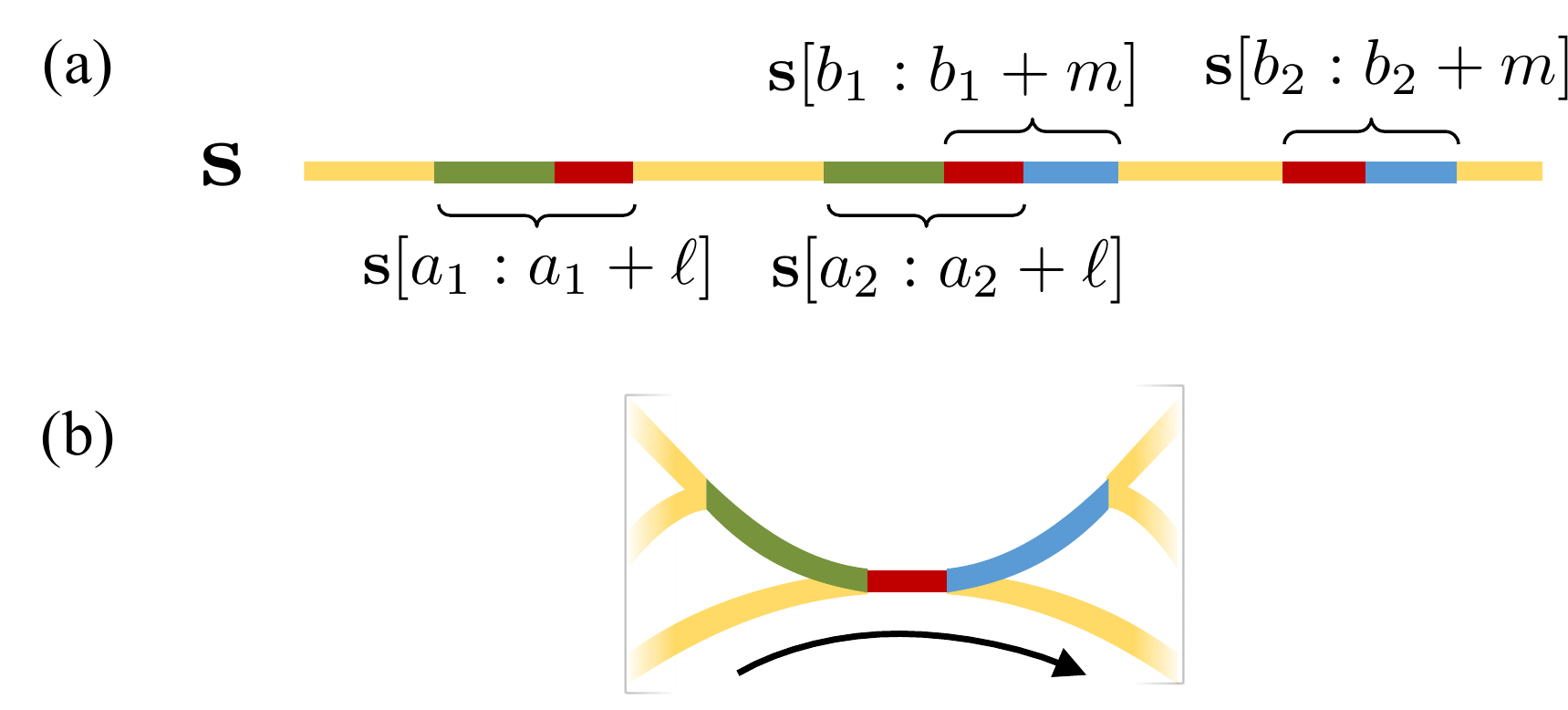}
\caption{(a) Illustration of linked repeats with link length $a_2 + \ell + 1-b_1$. (b) If we merge both repeats in the sequence graph, the link (red segment) creates a path that is not in the true sequence $\s$.} 
\label{fig:linked}
\vspace{-4mm}
\end{figure}
\fi

\iflong

We point out that the conditions in Theorem~\ref{thm:distortion} can be understood in light of the conditions for the standard greedy algorithm to achieve perfect assembly \cite{BBT}.
Notice that if all repeats in $\s$ are doubly bridged, the conditions in Theorem~\ref{thm:distortion} are satisfied for any $q$, implying that $D(G,\s) = 0$.
The standard greedy algorithm \cite{UkkonenGreedy}, on the other hand, achieves perfect assembly when all repeats are bridged, not necessarily doubly bridged \cite{BBT}.
Intuitively, the more stringent requirement of double bridging is the price paid to obtain guarantees in a range of $L$ where the genome is much more repetitive.

\fi

\section{Distortion on a Real Genome} \label{sec:numerical}


Clearly in practice we cannot verify whether the conditions in Theorems~\ref{thm:correctness} and \ref{thm:distortion} are satisfied, as we do not have access to the genome being sequenced.
The purpose of these results is to allow us to compute the rate-distortion tradeoff achieved by Algorithm \ref{alg:greedyalg} on previously assembled genomes.
This provides a framework to analyze the algorithm's performance and compare it to the fundamental lower bound (or to other algorithms).

\iflong
One can compute the probability that a given segment of length $\ell$ is not bridged by any of the $N$ length-$L$ reads as
\aln{
\left( 1 - \frac{L-\ell}{G} \right)^N \approx e^{-\frac{N}{G}(L - \ell)} \defi p_\ell,
}
and assume the bridging event to be independent for distinct segments in $\s$.
\fi
\iflong
For an organism whose whole genome $\s$ has been previously sequenced, we can compute the distribution of the length of the triple repeats in $\s$, as shown in Fig.~\ref{fig:saureus_sufficient}(a) for \mbox{\emph{S. aureus}}.
Given the list or triple repeat lengths $\T_\s$, we can then bound the probability that there is a triple repeat that is not all-bridged nor all-unbridged via the union bound as
\al{ \label{proberror1}
P_{{\rm triple}}(N,L) \defi \sum_{\ell \in \T_\s} {3\choose1} p_\ell (1-p_\ell)^2 +  {3\choose2} p_\ell^2 (1-p_\ell).
}
For a given target error probability $\ep$, we can then numerically compute the number of reads $N$ required to guarantee that the condition in Theorem \ref{thm:correctness} holds with probability at least $1-\ep$.
\fi
This yields the curve in Fig.~\ref{fig:saureus_sufficient}(b).

Similarly, by identifying the distribution of repeat lengths and characterizing which pairs of repeats are linked, one can compute the probability that conditions (b) and (c) in Theorem~\ref{thm:distortion} are not satisfied for a given $q$.
\iflong
This yields a second error probability $P_{q}(N,L)$.
By Theorem~\ref{thm:distortion}, it follows that the distortion achieved by the graph constructed via Algorithm \ref{alg:greedyalg} satisfies $D(G,\s) \leq D_q(\s)$ with probability at least $1-P_{{\rm triple}}(N,L) - P_q(N,L)$.
\fi
\iflong
\begin{figure}[h]
\centering
\includegraphics[width = 0.5\linewidth]{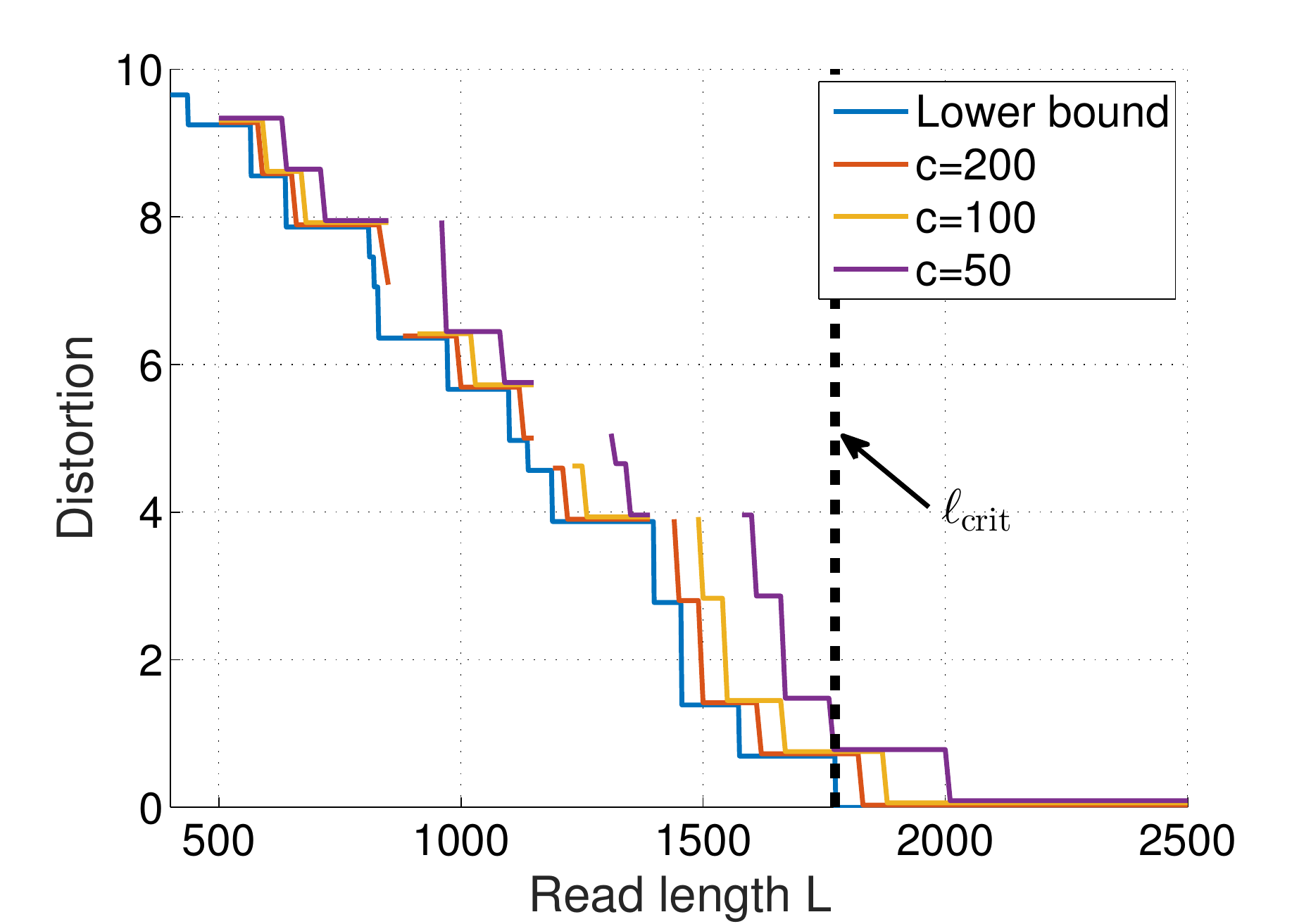}
\caption{Distortion achieved by Algorithm \ref{alg:greedyalg} with $k=300$ on \emph{S. aureus} with probability $0.99$ for different coverage depths $c = NL/G$, compared to the lower bound $D_L(\s)$. Gaps indicate that the probability of the conditions of Theorem \ref{thm:distortion} not being satisfied is at least $0.01$.} 
\label{fig:saureus_distortion}
\end{figure}
\fi
By computing $D_q(\s)$ for a range of values of $q$, which can be done using the well-known BEST Theorem\iflong \cite{BEST1,BEST2}, as described in Section \ref{sec:best}, \fi we can upper bound the distortion achieved by Algorithm \ref{alg:greedyalg} with a desired probability $1-\ep$.
Notice that $D_L(\s)$ is 
also the minimum distortion that can be achieved with reads of length $L$, which 
  provides a lower bound to the distortion that can be achieved by any algorithm.
In Fig.~\ref{fig:saureus_distortion} we show these curves computed for \emph{S. aureus} for different values of the coverage depth $c = NL/G$.
We notice that the upper bound curves follow the lower bound closely but have gaps in them, representing the ranges of $L$ where the conditions of Theorem~\ref{thm:correctness} are not satisfied with the desired probability, and the achieved distortion jumps to $D_1(\s)$.

\section{Proofs of Main Results}

\iflong


\subsection{Theorem \ref{thm:correctness}} \label{sec:lemproof}


Consider the position of each read in $\s$.
This uniquely defines an ordering on the reads, say $\r_1,\r_2,...,\r_n$, where $\r_1$ 
is chosen arbitrarily.  Note that some reads may map to multiple places in the sequence $\s$, in 
which case the read will appear multiple times in the ordering. 
We prove Theorem \ref{thm:correctness} via the following lemma.


\begin{lemma} \label{lem:overlaplem}
Assume the conditions in Theorem~\ref{thm:correctness} are satisfied. 
Suppose that reads $\r_i$ and $\r_j$ share a maximal substring $\x$ of length $\ell \geq k$, say, $\r_i [a:a+\ell-1] = \r_j [b:b+\ell-1] = \x$.
At the end of iteration $L-\ell$ of Algorithm \ref{alg:greedyalg} (which identifies matches of length $\ell$),
the paths corresponding to $\r_i [a:a+\ell-1]$ and $\r_j [b:b+\ell-1]$ are merged in the sequence graph if
\begin{enumerate}[label=(\roman*)]
\item $\r_i [a:a+\ell-1]$ and $\r_j [b:b+\ell-1]$ map to the same segment in $\s$, or
\item $\r_i [a:a+\ell-1]$ and $\r_j [b:b+\ell-1]$ map to unbridged repeats in $\s$.
\end{enumerate}
\end{lemma}


Given Lemma \ref{lem:overlaplem}, Theorem  \ref{thm:correctness} follows immediately.
We simply notice that from condition (i), at the end of the algorithm, the overlapping part of reads $\r_i$ and $\r_{i+1}$ (which must be of length at least $k$ when $\Rs$ $k$-covers $\s$) must be 
merged in the sequence graph.
Hence, by following the path corresponding to reads $\r_1,\r_2,...,\r_n$ in order, we spell out the sequence $\s$ and traverse every edge on the resulting sequence graph.


\begin{proof}[Proof of Lemma \ref{lem:overlaplem}]
We prove this by induction on $\ell = L, L-1, L-2,...,k$. 
Without loss of generality, we assume that
$i < j $.
%
When $\ell=L$, the Algorithm \ref{alg:greedyalg} just merges all repeated reads.
For  $\ell=L-1$, we have that if a pair of reads $\r_i$ and $\r_{j}$ 
have a matching substring of size $L-1$, they must correspond to a graph suffix or prefix in the 
beginning of the algorithm and will thus be merged 
(notice that in line \ref{line:merge} of Algorithm \ref{alg:greedyalg}, all instances of $\x \in X$ are merged).

Next, let us assume that the induction hypothesis holds up to the iteration where the 
algorithm searches for matches of length $\ell+1$ (iteration $L-\ell-1$ of the algorithm), 
and consider the iteration 
of the algorithm that searches for $\ell$-matches.
We consider cases (i) and (ii) separately.

 \textbf{Case (i):}
 In this case, as illustrated in Fig.~\ref{fig:overlapcase}(a), we must have $a = L-\ell+1$ and $b = 1$; i.e., the $\ell$-suffix of $\r_i$ matches the $\ell$-prefix of $\r_j$.
 Hence, if Algorithm \ref{alg:greedyalg} does not  merge $\r_i [L-\ell+1:L]$ and $\r_j [1:1+\ell-1]$ when it looks for matches of length $\ell$, 
 it must be the case that a longer suffix of $\r_i$ and a longer prefix of $\r_j$ were merged to other reads, say ${\bf t}_i$ and ${\bf t}_j$ in a previous iteration, as illustrated in Fig.~\ref{fig:overlapcase}(b), and both ${\bf t}_i$ and ${\bf t}_j$ have $\x$ as a substring.
If ${\bf t}_i$ is mapped to the same copy of $\x$ in $\s$ as $\r_i$, ${\bf t}_i$ must have an overlap with both $\r_i$ and $\r_j$ strictly greater than $\ell$, and by the induction hypothesis it would have been merged to both in previous iterations, causing $\r_i [L-\ell+1:L]$ and $\r_j [1:\ell]$ to also be merged.
Similarly, If ${\bf t}_i$ is mapped to the same copy of $\x$ in $\s$ as $\r_j$, $\r_i [L-\ell+1:L]$ and $\r_j [1:\ell]$ would be consequently merged.
Moreover, if ${\bf t}_i$ and ${\bf t}_j$ map to the same copy of $\x$ in $\s$ (not necessarily the same one as $\r_i$ and $\r_j$), they would have an overlap strictly greater than $\ell$, and by the induction hypothesis would be merged to each other, in turn causing $\r_i [L-\ell+1:L]$ and $\r_j [1:\ell]$ to be merged.


%
 \iflong
  \begin{figure}[t]
\centering
\includegraphics[width = 0.5\linewidth]{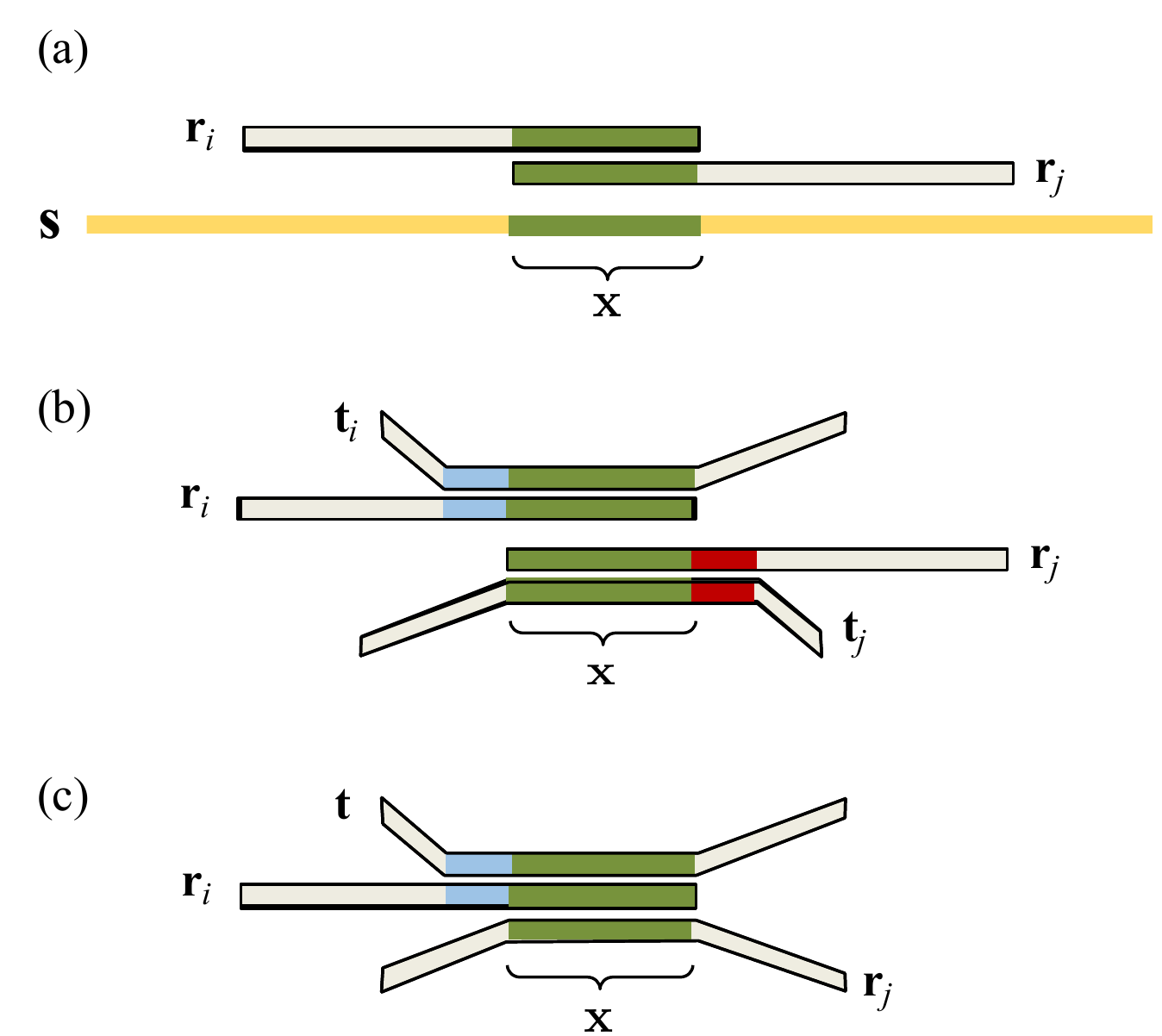}
\caption{(a) In case (i), the maximal match between $\r_i$ and $\r_j$ can be assumed to be an $\ell$-suffix of $\r_i$ and an $\ell$-prefix of $\r_j$; (b) In case (i), if at iteration $L-\ell$ the $\ell$-suffix of $\r_i$ and the $\ell$-prefix of $\r_j$ are not merged, $\r_i$ must have a longer suffix that was previously merged to a read ${\bf t}_i$ that extends $\r_i$ to the right, and $\r_j$ must have a longer prefix that was previously merged to a read ${\bf t}_j$ that extends $\r_j$ to the left; (c) In case (ii), if at iteration $L-\ell$ the $\ell$-suffix of $\r_i$ and the $\ell$-prefix of $\r_j$ are not merged, $\r_i$ must have a longer suffix that was previously merged to a read ${\bf t}i$ that extends $\r_i$ to the right.} 
\label{fig:overlapcase}
\end{figure}
 \else
   \begin{figure}[t]
\centering
\includegraphics[width = 0.8\linewidth]{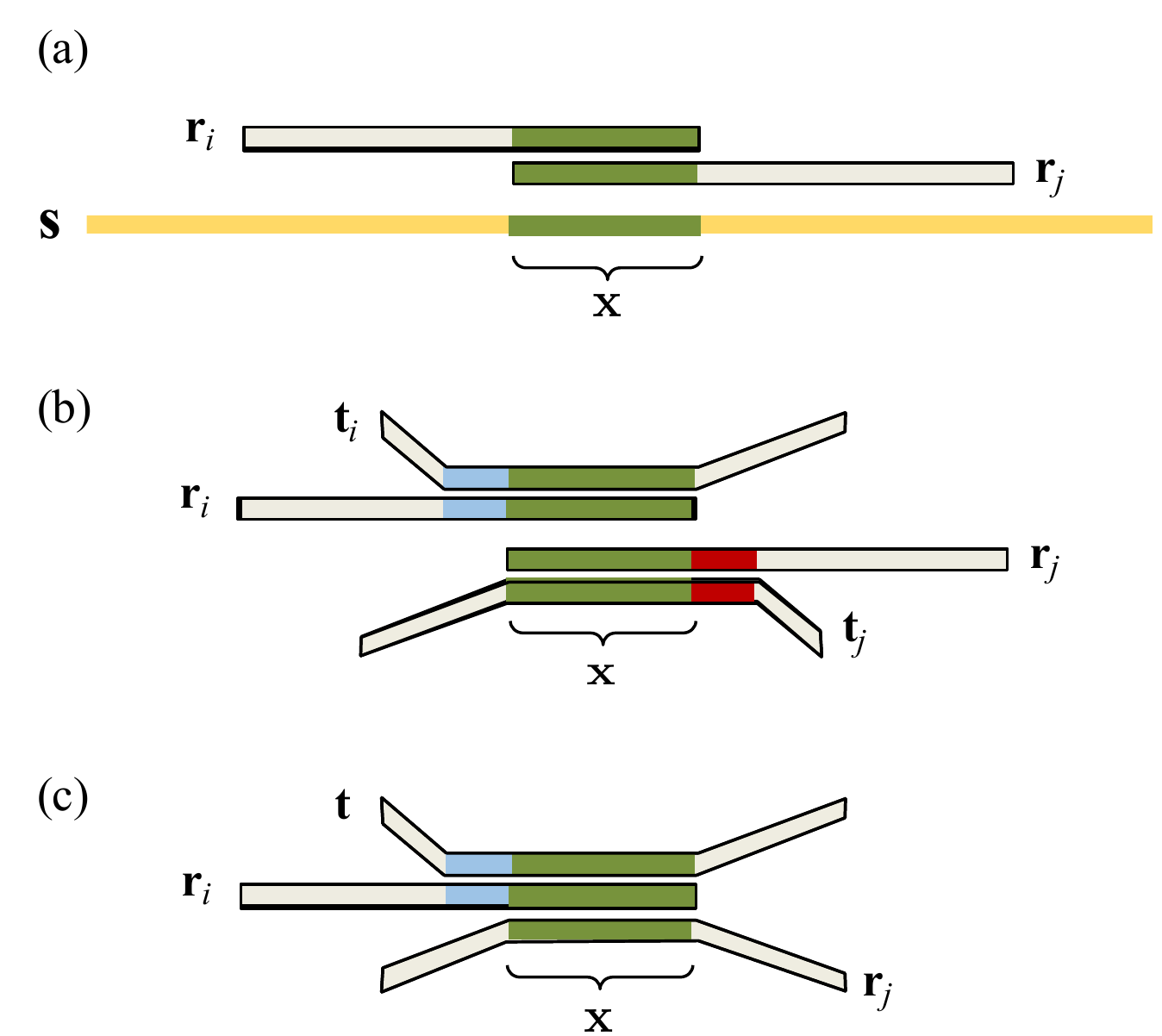}
\caption{(a) In case (i), the maximal match between $\r_i$ and $\r_j$ can be assumed to be an $\ell$-suffix of $\r_i$ and an $\ell$-prefix of $\r_j$; (b) In case (i), if at iteration $L-\ell$ the $\ell$-suffix of $\r_i$ and the $\ell$-prefix of $\r_j$ are not merged, $\r_i$ must have a longer suffix that was previously merged to a read ${\bf t}_i$ that extends $\r_i$ to the right, and $\r_j$ must have a longer prefix that was previously merged to a read ${\bf t}_j$ that extends $\r_j$ to the left; (c) In case (ii), if at iteration $L-\ell$ the $\ell$-suffix of $\r_i$ and the $\ell$-prefix of $\r_j$ are not merged, $\r_i$ must have a longer suffix that was previously merged to a read ${\bf t}i$ that extends $\r_i$ to the right.} 
\label{fig:overlapcase}
\end{figure}
 \fi

  
  The only case left to consider is when ${\bf t}_i$ and ${\bf t}_j$ map to two other copies of $\x$, implying that $\x$ is part of a triple repeat.
  Let $\y$ be the corresponding maximal triple repeat (of which $\x$ is a substring).
  Let $\r_i \rightarrow \r_j$ be the sequence obtained by concatenating $\r_i$ and $\r_j$; i.e., $\r_i [1:L-\ell] \concat \r_j $. 
  We first note that if
  the copy of  $\y$ in $\r_i \rightarrow \r_j$ was bridged, it would have to be bridged by a read
  $\r_k$, $ i < k < j$, which would have an overlap strictly longer than $\ell$ with both $\r_i$ and $\r_j$ and, by the induction hypothesis,
  would cause the entire segment to be merged.
  If the copy of $\y$ in $\r_i \rightarrow \r_j$ is not bridged, by the Theorem assumptions, $\y$ must be an all-unbridged
  triple repeat. 
This implies that the other two copies of $\y$, where ${\bf t}_i$ and ${\bf t}_j$ lie must be an unbridged repeat.
Since ${\bf t}_i$ and ${\bf t}_j$ must have a match strictly greater than $\ell$, 
by the induction hypothesis, ${\bf t}_i$ and ${\bf t}_j$ would be merged, causing $\r_i [L-\ell+1:L]$ and $\r_j [1:\ell]$ to be merged.

\textbf{Case (ii):}
First we notice that the maximal match between $\r_i$ and $\r_j$ cannot be internal on both reads, since that would imply a bridged repeat.
Thus, we can assume wlog that the match between $\r_i$ and $\r_i$ is a suffix at $\r_i$, but it need not be a prefix at $\r_j$; i.e., we may have $b > 1$.
  We note that, by the assumption in (ii), $\x$ is 
   part of an unbridged repeat in $\s$.
  If the algorithm does not merge them at the iteration that looks for $\ell$-matches,
a longer suffix of $\r_i$ must have beem merged in a previous iteration to a read ${\bf t}$ that extends to the right of $\x$,
as illustrated in Fig.~\ref{fig:overlapcase}(c).
Notice that ${\bf t}$ must also contain the substring $\x$.
We consider two cases.
First, if the substring $\x$ of ${\bf t}$ maps to the same place as $\r_i$ or to the same place as $\r_j$ in $\s$, then ${\bf t}$ and $\r_j$ must share a substring strictly longer than $\x$.
By the induction hypothesis, ${\bf t}$ would have been previously merged to $\r_j$, implying that $\r_i [L-\ell+1:L]$ and $\r_j [a:a+\ell-1]$ are also merged.

If the substring $\x$ of ${\bf t}$ does not map to the same place as $\r_i$ or $\r_j$, $\x$ must be part of a triple repeat in $\s$.
As we did in case (i), we let $\y$ be the maximal triple repeat containing $\x$, and we first note that if
  the copy of  $\y$ at $\r_i$'s location in $\s$ was bridged, it would have to be bridged by a read
  $\r_k$ which would have an overlap strictly longer than $\ell$ with both $\r_i$ and $\r_j$ and, by the induction hypothesis,
  would cause $\r_i$ and $\r_j$ to be merged.
  If the copy of $\y$ in $\r_i$ is not bridged, by assumption, $\y$ must be an all-unbridged
  triple repeat. 
This implies that the copies of $\y$ where ${\bf t}$ and $\r_j$ lie must be an unbridged repeat.
Since ${\bf t}$ and $\r_j$ must have a match strictly longer than $\ell$, by
the induction hypothesis they are already merged, and so are $\r_i [L-\ell+1:L]$ and $\r_j [a:a+\ell-1]$.
\end{proof}

\subsection{Theorem 2}\label{sec:thmproof}

We prove this result in three steps: 
\begin{itemize}
 \item First we define a sufficient graph $\G_{\U(\Rs)}$ by taking a cycle graph representation of $\s$ and merging the repeats that are not doubly-bridged by $\Rs$.
 \item Then we show that the distortion achieved by Algorithm \ref{alg:greedyalg} is at
 most $D(\G_{\U(\Rs)})$.
 \item To conclude the proof, we will show that when  the conditions in Theorem \ref{thm:distortion} are met, any path of $q-k$ edges in $\G_{\U(\Rs)}$ corresponds to a $q$-mer from $\C_q(\s)$, which implies that any 
Eulerian cycle in $\G_{\U(\Rs)}[\s]$ must define a sequence $\s'$ with $\C_q(\s) = \C_q(\s')$, which implies that $D(\G_{\U(\Rs)},\s) \leq D_q(\s)$.
\end{itemize}

In the context of a sequence graph of order $k$, the circular genome $\s$ can be represented as a 
cycle graph $\G = (\V,\E,\ch)$ with nodes $\V = \{1,...,G-k+1\}$ and edges 
$\E = \{(1,2),...,(G-k,G-k+1),(G-k+1,1)\}$ and $\ch(i,i+1) = \s[i:i+k-1]$.
We call this graph, illustrated in Fig.~\ref{fig:circle}(a), the cycle sequence graph of order $k$. 
We remark that  the cycle sequence graph is sufficient and achieves 
$D(\G,\s) = 0$, although we point out that other assembly graphs may achieve zero 
distortion as well.

A repeat of length $\ell$ in $\s$ corresponds to two $(\ell-k+1)$-node paths
on $\G$, say $[a : a+\ell-k]$ and $[b : b+\ell-k]$.
One can define a new sequence graph by taking $\G$ and contracting the paths
$[a : a+\ell-k]$ and $[b : b+\ell-k]$; i.e., merging the corresponding nodes and edges.
Notice that since the edges 
$(a,a+1),...,(a+\ell-k-1,a+\ell-k)$ and $(b,b+1),...,(b+\ell-k-1,b+\ell-k)$ correspond to the 
same $k$-mers, contracting the two paths is a well-defined operation.
In general, for a set of repeats 
\aln{
\T = {{\big\{}} & \left(\s[a_1 : a_1 + \ell_1], \s[b_1 : b_1 + \ell_1 ]\right), ...,  \\ 
& \left( \s[a_m : a_m + \ell_m], \s[b_m : b_m + \ell_m ] \right) {{\big\}}},
}
one can consider the sequence graph $\G_\T$ obtained by contracting 
the paths $[a_i: a_i+\ell_i-k]$ and $[b_i : b_i+\ell_i-k]$ corresponding to each repeat.
The resulting graph is illustrated in Fig.~\ref{fig:circle}(b).
It is straightforward to see that the order in which these merging operations are carried 
out does not affect the resulting graph.
\iflong
\begin{figure}[h]
\centering
\includegraphics[width = 0.5\linewidth]{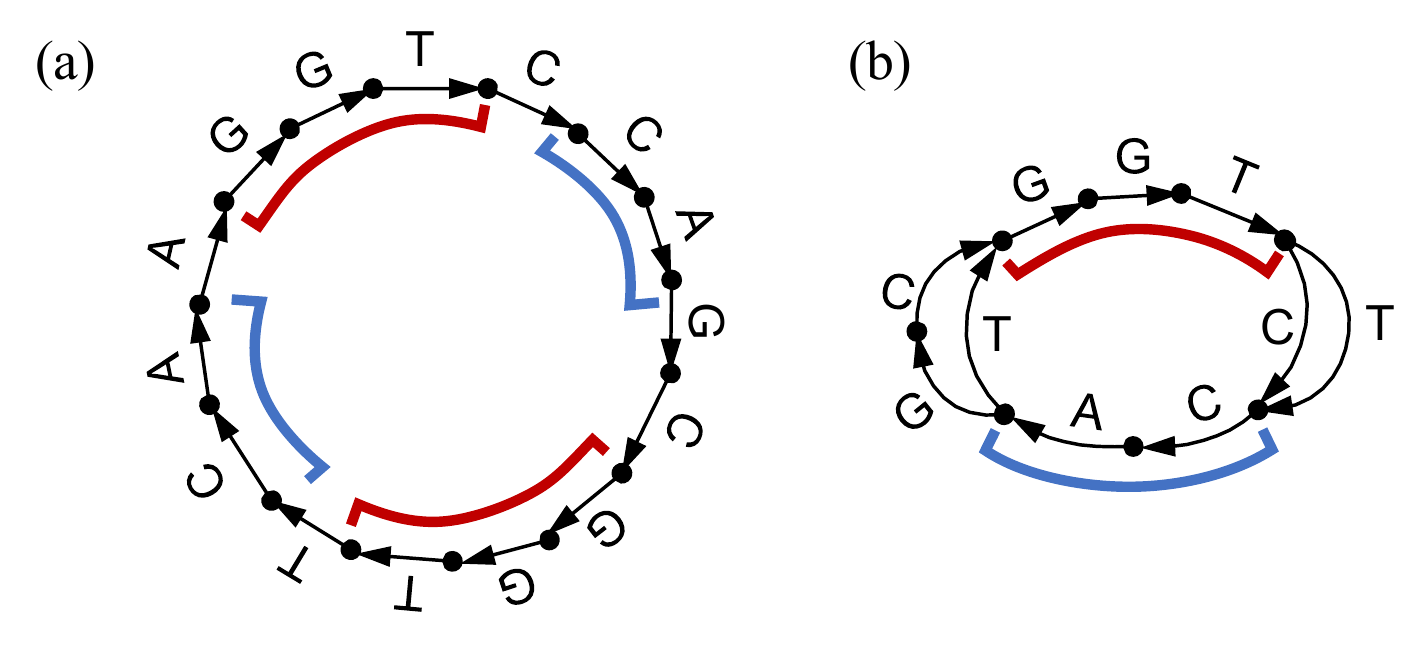}
\caption{(a) Cycle sequence graph $\G$ (of order $k=1$) for the sequence $\mathsf{G\,G\,T\,C\,C\,A\,G\,T\,C\,G\,G\,T\,T\,C\,A\,A}$; (b) Contracted graph $\G_\T$ where $\T$ corresponds to the two pairs of repeats shown in red and blue.} 
\label{fig:circle}
\end{figure}
\else
\begin{figure}[h]
\centering
\includegraphics[width = 0.8\linewidth]{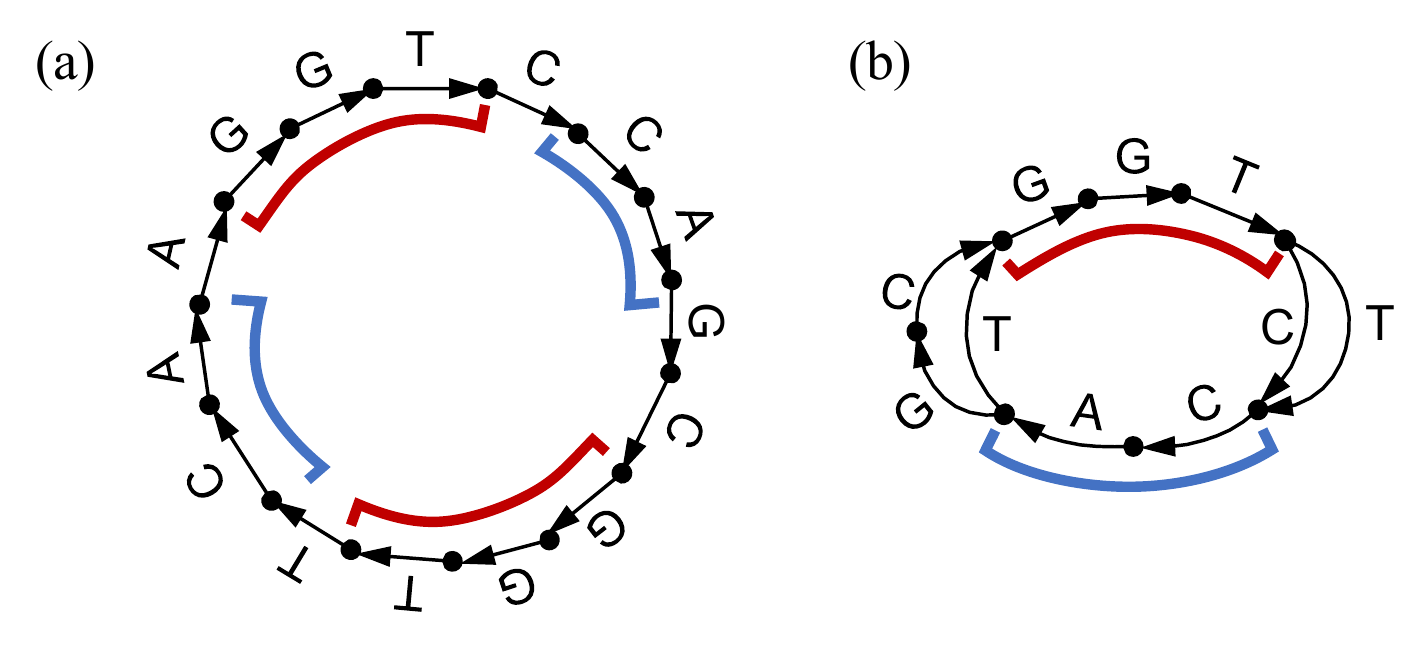}
\caption{(a) Cycle sequence graph $\G$ (of order $k=1$) for the sequence $\mathsf{G\,G\,T\,C\,C\,A\,G\,T\,C\,G\,G\,T\,T\,C\,A\,A}$; (b) Contracted graph $\G_\T$ where $\T$ corresponds to the two pairs of repeats shown in red and blue.} 
\label{fig:circle}
\end{figure}
\fi

Given the set of reads $\Rs$ from $\s$, one can then define the set $\U(\Rs)$ 
of maximal non-doubly-bridged repeats.
As described above, this set of repeats defines a sequence graph $\G_{\U(\Rs)}$ where each repeat in $\U(\Rs)$ 
is merged. 

Next, we show that the when the conditions in Theorem \ref{thm:distortion} are satisfied, the 
sequence graph produced by Algorithm \ref{alg:greedyalg} has a distortion that is 
upper bounded by $D\left(\G_{\U(\Rs)},\s\right)$. 
Consider the initial sequence graph in Algorithm \ref{alg:greedyalg} 
(illustrated in Fig.~\ref{fig:initial}).
Since each read corresponds to a path (or multiple paths) of the form $[a:a+L-k]$ in $\G$, we can consider labeling each node in the initial graph in Algorithm \ref{alg:greedyalg} with its corresponding node in $\V = \{1,...,G-k+1\}$ (or set of nodes, in case the read matches multiple segments of $\s$).
As the algorithm progresses, and nodes are merged, imagine that we take the union of the set of
numbers at the merged nodes.
If the conditions  of Theorem \ref{thm:correctness} are satisfied, the Chinese Postman cycle corresponding to $\s$ in the final sequence graph will be determined by following the numbers in 
$\V  = \{1,...,G-k+1\}$ .
We claim the following:

\begin{claim}\label{lem:bridgerep}
Suppose the conditions in Theorem \ref{thm:distortion} are satisfied. 
If a merging operation of Algorithm~\ref{alg:greedyalg} merges two paths that do not correspond to the same segment of $\s$, the two paths must correspond to an unbridged repeat (or a substring of it) in $\s$.
\end{claim}

\begin{proof} 
Suppose that in some iteration, Algorithm~\ref{alg:greedyalg} merges the segment $\x$ of two reads $\r_i$ and $\r_j$ that do not truly overlap in $\s$.
From the way the algorithm chooses sequences $\x$ to merge, $\x$ can be assumed wlog to be the suffix of some read $\r_{\ell}$ (possibly $\r_i$ or $\r_j$), and no longer suffix of $\r_{\ell}$ should have been previously merged.
Since the segment $\x$ in $\r_i$ and $\r_j$ do not correspond to the same segment in $\s$, we can assume wlog that the segment $\x$ in $\r_{\ell}$ and $\r_i$ are not the same segment in $\s$.
Since no longer suffix of $\r_\ell$ has been previously merged, the copy of $\x$ at $\r_\ell$ is not bridged.
If the copy of $\x$ at $\r_j$ corresponds to the same $\x$ as $\r_\ell$, the repeat is not doubly-bridged and the claim follows.
If the copy of $\x$ at $\r_j$ is a third copy, we have a triple repeat.
By condition (b) in Theorem~\ref{thm:distortion}, this must be an all-unbridged triple repeat, and the claim follows.
\end{proof}

From Lemma \ref{lem:bridgerep}, at each merging operation of Algorithm \ref{alg:greedyalg},
we can only merge two paths that do not correspond to the same segment in $\s$ 
if they correspond to a repeat that is not doubly-bridged.
This gives us that every time we merge two nodes that have distinct labels 
(and we take the union of
the labels), we must be merging nodes that are merged in $\G_{\U(\Rs)}$.
Therefore, $\G_{\U(\Rs)}$ is either equal to the output of Algorithm \ref{alg:greedyalg} 
or can be obtained by performing further node contractions.
This means that any Eulerian cycle in the graph $G[\s]$ is also an Eulerian cycle in $\G_{\U(\Rs)}[\s]$,  giving us 
that the distortions of these two sequence graphs are related by 
\aln{
D(G,\s) \leq D(G_{\U(\Rs)},\s),
}
as claimed.


Next we  bound $D(G_{\U(\Rs)},\s)$.
Let $\din(v)$ and $\dout(v)$ correspond to the in-degree and out-degree of $v$ respectively.
First, we make the following claim:

\begin{claim}  \label{claim:qpath}
Any (directed) $(q-k)$-edge path in $G_{\U(\Rs)}$ starting at a node $v_1$ with
$\din(v_1) > 1$ and visiting nodes $v_2,v_3,...,v_{q-k+1}$ must have the property that 
$\dout(v_i) = 1$ for $i=1,...,q-k+1$.
\end{claim}

\begin{proof} First we notice that a node $v$ in $G_{\U(\Rs)}$ can only have $\din(v) > 1$ if it is the 
starting node of some repeat in $\U(\Rs)$.
Similarly, a node $v$ can only have $\dout(v) > 1$ if it is the ending node of some repeat 
in $\U(\Rs)$.
Now, suppose by contradiction that $\dout(v_i) > 1$ for some $i \leq q-k+1$, and let $m$ be the first such index.
Clearly, when any Chinese Postman cycle reaches $v_1$, it must traverse the edges in the path until $v_m$, and
the path $v_1,...,v_m$ must correspond to the beginning of the repeat started at $v_1$.
Since $\dout(v_m)>1$, $v_m$ corresponds to the end of a repeat.
Since $\U(\Rs)$ only contains repeats longer than $q$, $v_1$ and $v_m$ cannot correspond to the beginning and end of the same repeat.
This means that $\s$ has two linked repeats in $\U(\Rs)$ (hence not doubly bridged) with a link length 
$m+k-2 \leq q$ (notice that path $(v_1,...,v_m)$ corresponds to a string of length $m+k-2$). 
But this is a contradiction to the assumption that, for every pair of linked repeats with link 
length $\ell$ satisfying $k-1 \leq \ell \leq q$, at least one is doubly bridged.
\end{proof}

Next, we use Claim \ref{claim:qpath} to show that any Eulerian cycle in $G_{\U(\Rs)}[\s]$ must correspond to a sequence $\x$ with $\C_q(\x) = \C_q(\s)$,
which directly implies that 
\aln{
D(G_{\U(\Rs)},\s) \leq D_q(\s).
}
To see this, consider an arbitrary Eulerian cycle in $G_{\U(\Rs)}[\s]$
corresponding to some sequence  $\x$.
Any $q$-mer of $\x$ corresponds to 
a $(q-k)$-edge path $(v_1,...,v_{q-k+1})$ in $G_{\U(\Rs)}$.
Now, Claim \ref{claim:qpath} guarantees that the nodes in the path $(v_1,...,v_{q-k+1})$ with $\dout(v_i) > 1$ must precede the nodes with 
$\din(v_i) > 1$; i.e., there exists a $c \in \{1,...,q-k\}$ such that $\din(v_i) = 1$ for
$i=1,...,c$ and $\dout(v_i) = 1$ for $i=c+1,...,q-k+1$.
Since any Eulerian cycle must traverse edge $(v_c,v_{c+1})$, it must arrive there through the path $(v_1,...,v_c)$.
Since the cycle must continue after $v_{c+1}$, it needs to traverse the remainder of the path $(v_{c+1},...,v_{q-k+1})$.
Therefore the number of complete traversals of the path $(v_1,...,v_{q-k+1})$ by the Eulerian cycle is precisely the same as the multiplicity of edge $(v_c,v_{c+1})$ in $G_{\U(\Rs)}[\s]$.
Therefore, any $(q-k)$-edge path in $G_{\U(\Rs)}$ is traversed the same number of times by any Eulerian cycle in $G_{\U(\Rs)}[\s]$ and, in particular, by the Eulerian cycle corresponding to $\s$.
This means that $\C_q(\x) = \C_q(\s)$, as we intended to prove.

%
%

\subsection{BEST Theorem} \label{sec:best}


The BEST Theorem provides an efficient way to count the number of Eulerian cycles in a Eulerian directed graph.
%
 
\begin{theorem}[\cite{BEST1,BEST2}] \label{thm:best}
The number of Eulerian cycles in an Eulerian multigraph $G=(V, E)$, is given by
\begin{equation}
T_G \prod_{v \in V}  (\dout(v)-1)! ,
\end{equation}
where $T_G$ is the number of arborescences of $G$, and $\dout(v)$ is the out-degree of $v$.
\end{theorem}

Notice that, from the point of view of Theorem \ref{thm:best}, if an edge has a multiplicity higher than one, each copy is considered as a distinct
edge. 
But in order to compute $D_q(\s)$ as defined in (\ref{eq:kdist}) we need to correct for this double-counting, and a slightly modified version of the BEST Theorem has to be applied. 
For every pair of vertices  $ u,v \in V$, let $m(u,v)$ be the multiplicity of the 
edge $(u,v)$.
Thus
\begin{equation}
\sum_{v \in V} m(u,v) = \dout(u).
\end{equation}


\begin{cor}
Suppose an Eulerian multigraph $G = (V,E)$ contains an edge with multiplicity $1$.
The number of Eulerian cycles in $G$, distinct up to edge multiplicity, is given by
\begin{equation}
\ec(G) = \prod_{(u,v) \in V \times V}\frac{1}{m(u,v)! }T_G  \prod_{v \in V} (\dout(v_i)-1)! 
\end{equation}

\end{cor}

\begin{proof}
\Cref{thm:best} views Eulerian cycles that can be obtained from each other by reordering the traversals of an edge with multiplicity greater than one as distinct Eulerian cycles.
One can then define an equivalence relationship between Eulerian cycles where $c \sim c'$ if $c'$ can be obtained from $c$ by reordering the traversals of each edge.
Since $G$ has an edge of multiplicity one, it can be thought of as the first edge of an Eulerian cycle $c$.
Thus it is impossible that by reordering the traversals of each edge in $c$ we obtain $c$ itself.
This implies that each equivalence class has size exactly $\prod_{(u,v) \in V \times V}{m(u,v)! }$, and the result follows.
\end{proof}

\fi

{\footnotesize
\bibliographystyle{IEEEtran}
\bibliography{refsc}
}

\end{document}